quant-ph/0604093
\documentclass[pra,12pt,eqsecnum]{revtex4}
\usepackage{graphicx}

\begin{document}

\def\BE{\begin{equation}}
\def\EE{\end{equation}}
\def\BY{\begin{eqnarray}}
\def\BEA{\begin{eqnarray}}\def\EY{\end{eqnarray}}\def\EEA{\end{eqnarray}}
\def\q{\vec q}
\def\r{\vec r}
\def\L{\label}
\def\nn{\nonumber}
\def\({\left (}
\def\){\right)}
\def\[{\left [}
\def\]{\right]}
\def\o{\overline}
\def\BA{\begin{array}}
\def\EA{\end{array}}
 \def\ds{\displaystyle}

\title{Induced photon statistics in three-level lasers }
\author{T.~Golubeva and Yu.~Golubev}
\address{V. A. Fock Physics Institute, St.~Petersburg State University,\\
ul Ul'yanovskaya 1, 198504 St.~Petersburg, Stary Petershof,
Russia}
\date{\today}

\begin{abstract}
The statistical properties of three-level lasing are investigated
theoretically. It is assumed  that the three-level medium is
coherently excited by another laser with an arbitrary photon
statistics. It is proved that, under the specific conditions, the
photon statistics of the three-level laser duplicate the photon
statistics of the exciting laser. We call this phenomenon an {\it
induced photon statistics}. We suggest  to use this to analyze the
statistical properties of a laser involved into a feedback
process. Applying this laser for the coherent pump of a
three-level laser, we can follow  its photon statistics by means
of direct following the three-level generation. In accordance with
Ref.~\cite{Milburn}, we conclude that the feedback in itself is
unable to generate the non-classical manifestation    in the laser
field.

\end{abstract}

%PACS number(s): 42.55.Px, 42.60.Mi, 42.50.Dv

\maketitle

\section{Introduction}
The three-level laser is a unique source of a bright light, where
the quantum peculiarities in the field are produced automatically
without any supplementary efforts such as, for example, an
establishment of a regular pump Ref.~\cite{Golubev}. For this
source, the stabilizing mechanism was investigated in detail and
this is connected exclusively with the non-linear behaviour of the
three-level medium interacting with two coherent fields
Ref.~\cite{Ritsch}.

At the same time there is another field manifestation more
interesting for us here. In Ref.~\cite{Rost}, it was illustrated
that, in three-level lasing, the photon statistics can turn out to
be dependent on statistical properties of  the other laser that is
used for the coherent pump of the three-level medium. In the cited
work, it was suggested to employ this phenomenon for  an
improvement of the quantum statistical parameters in the
sub-Poissonian three-level generation.

We are going to consider some effect that  is inexplicitly already
contained in the formulas of Ref.~\cite{Rost} but was in no way
accented there. We shall demonstrate here that it is possible to
ensure the conditions when the photon statistics in three-level
lasing are not only dependent but duplicate the one in the pump
laser. We call this phenomenon  {\it induced photon statistics
({\it IPS})}.

To our mind, the {\it {\it IPS}} is of our interest in itself as a
fundamental. At the same time one can see, where this phenomenon
could be applied. Really, for example, if there is a situation
when the laser beam turns out to be inaccessible for direct
testing, then you could follow its statistical properties by means
of following three-level lasing. For that, you need only to apply
your laser for the coherent excitation of the last and ensure the
conditions for existing the {\it {\it IPS}}.

The laser beam can be inaccessible for the direct investigation by
the quite different reasons. For example, as is known some
frequency ranges are extremely hardly  detected or the light beam
from  the laser involved into the feedback process is inaccessible
too.

The case with the feedback is more interesting for us and we shall
touch it here. This problem was widely considered in different
scientific groups experimentally by Yammamoto, Fofanov, Mosalov
\cite{Yammamoto, Fofanov, Mosalov} and theoretically by Troshin,
Wiseman, and Milburn \cite{Milburn, Troshin}. Everybody of them
has fixed that the photocurrent noise from the laser in the {\it
feedback loop (FBL)} can be reduced even below the quantum limit.
This experimentally demonstrated effect has produced the
determined hopes that we are able to obtain sub-Poissonian lasing
by means of only the establishment of the acceptable feedback.

However there are reasons to think differently too. The point is
that the observed noise reduction was easy explained not only
within the framework of quantum electrodynamics but  as well as
the classical one. This can mean only that the
 observed experimentally reduction of the current noise in the
{\it FBL} is in no way connected logically with the photon noise
one. In order to make some conclusions relative to the photon flux
in this case, it is not quite enough to follow only the
photocurrent and some supplementary measuring procedures must be
provided.

From our standpoint, the determined clarity was reached  by
Weisman and Milburn in Ref.~\cite{Milburn}. There the authors
could insist that the feedback is unable  to ensure conditions for
the generation of the non-classical light. The conclusion was made
on the basis of the very formal considerations. They have proved
that if the internal dynamics of the cavity generate a light state
that has a positive Glauber-Sudarshan P-function, this positivity
survives under switching the feedback on.

We think that it would be useful to illustrate the same on the
less formal level by applying any measuring procedure, on the
basis of any mental experimental situation. One of the possible
procedure, to our mind, could be connected with the{\it IPS} and
we are going to consider this way below in this article.

This paper is organized as follows. In Sec.~\ref{II}, the
three-level laser model and the observed signals are considered.
In Sec.~\ref{III}, the quantum statistical theory of the joint
system consisting of  the three-level laser and the coherently
exciting laser is presented. In Sec.~\ref{IV}, the conditions for
existing the{\it induced photon statistics} are found. In
Sec.~\ref{V}, the phenomenon of the{\it induced photon statistics}
is applied for analysis of the laser radiation from laser in the
feedback loop.

\section{Joint system of  exciting and three-level lasers\L{II}}
We shall treat theoretically the same mental experiment as in
Ref.~\cite{Rost}.  One can see in Fig.~1 that there are two lasers
interacting with each other by means of their common intracavity
space. One of them  is a two-level
 laser that is designated in the figure  as '2-laser' (after {\it
 (2L)}).  Its intracavity field is used for a coherent pump of
 a three-level laser ('3-laser' in the figure
 or simply {\it (3L)}).

 The feature of our theoretical approach to this
 experimental situation  is of the same as in Ref.~\cite{Rost}.
 Usually, in the similar configurations,  it is assumed that
 although  the three-level medium is excited by the two-laser
 intracavity   field, nevertheless this interaction does not
 affect on lasing itself. We are not going to be restricted
  by this assumption. On the contrary, the case when the main losses in the {\it
 (2L)} cavity are connected just with the three-level medium
 excitation is of our main interest here.

In Fig.\ 2, both  two- and three-level atomic configurations are
shown together. We assume that the two-level medium (on the left
in the figure) is pumped incoherently  to the upper atomic state
$|1\rangle$ with the mean rate $R$ and there are  spontaneous
decays of the atomic states $|1\rangle$ and $|2\rangle$
correspondingly  with the rates $\gamma_1$ and $\gamma_2$.

It is supposed that three-level lasing takes place on the
transition $|\tilde1\rangle\leftrightarrow|\tilde2\rangle$ (on the
right in the figure). In order to ensure the {\it (3L)}
generation, the three-level medium is excited   coherently on the
transition $|\tilde3\rangle\to|\tilde1\rangle$ by a radiation from
the {\it (2L)} and incoherently on the transition
$|\tilde2\rangle\to|\tilde3\rangle$ with the rate $
\tilde\gamma_2$. Besides there is a spontaneous decay
$|\tilde1\rangle\to|\tilde2\rangle$ with the rate $
\tilde\gamma_1$. Hereinafter, the sign 'tilde' over any symbol
denotes that this symbol belongs to  the {\it (3L)}.

 Let the output signals in our mental experiment be
the photocurrent spectra
 $(\delta i^2)_\omega$ and $(\delta\tilde i^2)_\omega$ correspondingly
 for a registration of  the {\it (2L)} and {\it (3L)} radiation.
 These spectra are defined as factors near the
delta-functions in the pair correlation functions:
\BY
&&\o{\delta i_\omega\;\delta i_{\omega^\prime}}=(\delta
i^2)_\omega\;\delta(\omega+\omega^\prime),\qquad\o{\delta \tilde
i_\omega\;\delta \tilde i_{\omega^\prime}}=(\delta\tilde
i^2)_\omega\;\delta(\omega+\omega^\prime).\L{2.1}
\EY
The spectral components $\delta i_\omega$ and the photocurrent
fluctuation $\delta i(t)=i(t)-\o i$ are coupled by the Fourier
transformation:
\BY
&&\delta
i_\omega=\frac{1}{\sqrt{2\pi}}\int\limits_{-\infty}^{+\infty}\delta
i(t)\;e^{i\omega t}dt,\qquad\delta
i(t)=\frac{1}{\sqrt{2\pi}}\int\limits_{-\infty}^{+\infty}\delta
i_\omega\;e^{-i\omega t}d\omega\L{2.2}
\EY
and similarly for the other variable $\delta\tilde i_\omega$.

\section{Theory of the joint system of two lasers\L{III}}

\subsection{Master equation}
In the work \cite{Rost}, the  system of two lasers  was considered
within the framework of  the master equation theory. The master
equation was constructed in the Glauber diagonal representation in
the limit of the high-Q cavity and the small photon fluctuations.
the obtained master equation reads:
\BY
&& \frac{\partial R(\varepsilon,\tilde\varepsilon,t)}{\partial
t}=\frac{\partial}{\partial
\varepsilon}\((\kappa+\kappa_0)\varepsilon-\tilde\kappa\tilde\varepsilon\)R+
\frac{\partial}{\partial
\tilde\varepsilon}\(2\tilde\kappa\tilde\varepsilon-\kappa_0\varepsilon\)R+\nn\\
&&+(\kappa+\kappa_0)n\xi\frac{\partial^2 R}{\partial
\varepsilon^2}-\tilde\kappa\tilde n\frac{\partial^2 R}{\partial
\tilde\varepsilon^2}+\tilde\kappa\tilde n\frac{\partial^2
R}{\partial \varepsilon\partial
\tilde\varepsilon}+\{\cdots\}.\L{3.1}
\EY
As is known, the Glauber field amplitude
$\alpha=\sqrt{u}\;\exp(i\varphi)$ and $\tilde\alpha=\sqrt{\tilde
u}\;\exp(i\tilde\varphi) $ are introduced as eigen-numbers of the
corresponding annihilation operators:
\BY
&&\hat a|\alpha\rangle=\alpha|\alpha\rangle,\qquad\hat {\tilde
a}|\tilde\alpha\rangle=\tilde\alpha|\tilde\alpha\rangle,\qquad\[\hat
a,\hat a^\dag\]=1,\qquad\[\hat {\tilde a},\hat {\tilde
a}^\dag\]=1.
\EY
The limit of the small photon fluctuations  means that
$\varepsilon=u-n$ and $\tilde\varepsilon=\tilde u-\tilde n$ turn
out to be much less than the corresponding semi-classical meanings
$n,\tilde n$:
\BY
&&\varepsilon\ll n,\qquad\tilde\varepsilon\ll\tilde n.
\EY
The photon matrix $R(\varepsilon,\tilde\varepsilon,t)$ is
introduced by the phase integrating  the Glauber
P-quasi-distribution:
\BY
&& R(\varepsilon,\tilde\varepsilon,t)=\int\!\!\int d\varphi\;
d\tilde\varphi\; P(\alpha,\tilde\alpha,t).
\EY
The frequency parameters $\kappa$ and $\tilde\kappa$ are the
spectral mode widths for the {\it (2L)} and the {\it (3L)};
\BY
\kappa_0=\tilde\gamma_2\;\(\tilde g_{13}/\tilde g_{12}\)^2\;\tilde
N/\tilde n
\EY
is the mean rate of the coherent excitation of the three-level
medium $\tilde g_{13}$ and $\tilde g_{12}$ are the coupling
constants for the dipol interaction  of the field with the
three-level atom on the transitions $(1-3)$ and $(1-2)$, $\tilde
N$ is the three-level atom number.

A statistical aspect of the {\it {2L}} is determined by the
arbitrary Mandel parameter $\xi$ that depends on statistical
features of the pump of the laser medium. The last is determined
by only parameter  $p=-2\xi$ ($p<1$). By varying this parameter we
can obtain the Poissonian ($p=0$), sub-Poissonian ($0<p<1$)  or
super-Poissonian ($p<0$) laser.

As for semiclassical solutions $n$ and $\tilde n$, the following
useful equalities take place for them:
\BY
&&\tilde\kappa \tilde n=\kappa_0 n\qquad
(\kappa+\kappa_0)n=R.\L{3.5}
\EY
In order to simplify our formal expressions we have written all of
them in the saturation regime relative to both lasers and taken
the condition $\tilde\gamma_2\ll\tilde\gamma_1$. Also, it was
assumed that there are no any non-linear manifestations in the
coherent excitation of the three-level medium.

In the master equation, the notation $\{\cdots\}$ informs us that
as a matter of fact there are the other terms containing the
higher order derivatives with respect to $\varepsilon$ and
$\tilde\varepsilon$. As is known we have no any possibility to
avoid writing these terms in the case of the non-classical fields.
However for the description of correlation experiments of the
lowest orders, these terms nothing at all introduce.

\subsection{Langevin equations}
Let us rewrite our master equation theory in the terms of the
corresponding differential Langevin equations for the variables
$\varepsilon $ and $\tilde\varepsilon $:
\BY
&&\dot\varepsilon=-(\kappa+\kappa_0)\varepsilon+\tilde\kappa\tilde\varepsilon+F(t),\L{3.6}\\
&&\dot{\tilde\varepsilon}=-2\tilde\kappa\tilde\varepsilon+\kappa_0\varepsilon+\tilde
F(t).\L{3.7}
\EY
The pair correlation functions for the stochastic sources $F$ and
$\tilde F$ are given by:
\BY
&&
\o{F(t)F(t^\prime)}=2\xi(\kappa+\kappa_0)n\;\delta(t-t^\prime),\\
&&\o{\tilde F(t)\tilde F(t^\prime)}=-2\tilde\kappa\tilde
n\;\delta(t-t^\prime),\\
&&\o{F(t)\tilde F(t^\prime)}=\tilde\kappa\tilde
n\;\delta(t-t^\prime).
\EY
The more complicated correlations are connected with terms
$\{\cdots\}$ in master equation (\ref{3.1}) and are not essential,
  because, in our consideration here, we are going  to be restricted
only by the pair correlations for the photon variables and we do
not need to know the correlation function of the higher orders for
stochastic sources.

In the c-number theory, the intracavity photon fluctuations are
easy coupled with the corresponding photocurrent fluctuations $
\delta i$ and $ \delta \tilde i$ by means of the following
equalities:
\BY
&&\delta i=\kappa\varepsilon+S(t),\qquad\o{S(t)S(t^\prime)}=\o i
\;\delta(t-t^\prime),\qquad \o i=\kappa n,\\
&&\delta \tilde i=\tilde\kappa\tilde\varepsilon+\tilde
S(t),\qquad\o{\tilde S(t)\tilde S(t^\prime)}=\o{\tilde i}
\;\delta(t-t^\prime),\qquad \o{\tilde i}=\tilde\kappa\tilde n.
\EY
The stochastic sources here reflect the random process of the
light absorption under detecting.

As was mentioned above, we are going to consider the photocurrent
spectra as the signal in our mental experiment. One can see that
the spectral components of the currents
\BY
&&\delta
i_{\omega}=\kappa\varepsilon_{\omega}+S_{\omega},\L{3.13}\\
&&\delta \tilde
i_{\omega}=\tilde\kappa\tilde\varepsilon_{\omega}+\tilde
S_{\omega}\L{3.14}
\EY
are coupled directly with the spectral components
$\varepsilon_{\omega}$ and $\tilde\varepsilon_{\omega}$. Thus it
is better to rewrite  equations (\ref{3.6}) and (\ref{3.7}) in the
Fourier domain:
\BY
&&-i\omega\varepsilon_{\omega}=-(\kappa+\kappa_0)\varepsilon_{\omega}
+\tilde\kappa\tilde\varepsilon_{\omega}+F_{\omega},\L{3.15}\\
&&-i\omega\tilde\varepsilon_{\omega}=-2\tilde\kappa\tilde\varepsilon_{\omega}
+\kappa_0\varepsilon_{\omega}+\tilde F_{\omega}.\L{3.16}
\EY
The corresponding pair correlation functions for the spectral
stochastic sources read:
 \BY
&& \o{F_\omega F_{\omega^\prime}}=2\xi(\kappa+\kappa_0) n\;
\delta(\omega+\omega^\prime),\qquad\o{\tilde F_{\omega}\tilde
F_{\omega^\prime}}=-2\tilde\kappa\tilde
n\;\delta(\omega+\omega^\prime),\nn\\
&&\o{ F_{\omega}\tilde F_{\omega^\prime}}=\tilde\kappa\tilde n
\;\delta(\omega+\omega^\prime),\L{3.19}\\
&& \o{S_\omega S_{\omega^\prime}}=\kappa n\;
\delta(\omega+\omega^\prime),\qquad\o{\tilde S_{\omega}\tilde
S_{\omega^\prime}}=\tilde\kappa\tilde n
\;\delta(\omega+\omega^\prime).\L{3.21}
\EY

\section{Photocurrent spectra\L{IV}}
\subsection{General conditions}
From algebraic Eqs.~(\ref{3.15}) and (\ref{3.16}) it is not
difficult to express the photon fluctuations via the stochastic
sources in the following form:
\BY
&&\varepsilon_\omega=\frac{(2\tilde\kappa-i\omega)F_\omega+\tilde\kappa\tilde
F_\omega}{(\kappa+\kappa_0-i\omega)(2\tilde\kappa-i\omega)-\kappa_0\tilde\kappa},\\
&&\tilde\varepsilon_\omega=\frac{\kappa_0F_\omega+(\kappa+\kappa_0-i\omega)\tilde
F_\omega}{(\kappa+\kappa_0-i\omega)(2\tilde\kappa-i\omega)-\kappa_0\tilde\kappa}.
\EY
Now we have a possibility to calculate in the explicit form the
photon  spectral densities $(\varepsilon^2)_\omega$ and $
(\tilde\varepsilon^2)_\omega$, i. e., the photocurrent spectra
too. According to  (\ref{3.13}) and (\ref{3.14}) the last are
expressed via the first by:
\BY
&&(\delta \tilde
i^2)_\omega=\tilde\kappa^2(\tilde\varepsilon^2)_\omega+(\tilde
S^2)_\omega,\\
&&(\delta i^2)_\omega=\kappa^2(\varepsilon^2)_\omega+(S^2)_\omega.
\EY
After the non-complicated algebraic manipulations, one can obtain
the photocurrent in the {\it (3L)}-channel in the form:
\BY
&&(\delta \tilde i^2)_\omega/\o {\tilde
i}=1-2\tilde\kappa^2\frac{\omega^2+(\kappa+\kappa_0)(\kappa-\kappa_0\xi)
}{\(\omega^2-\tilde\kappa(2\kappa+\kappa_0)
\)^2+\omega^2\(2\tilde\kappa+\kappa+\kappa_0\)^2 },\L{4.5}
\EY
and in the {\it (2L)}-channel:
\BY
&& (\delta i^2)_\omega/\o {
i}=1+2\kappa\frac{\kappa_0\tilde\kappa^2+\xi(\kappa+\kappa_0)(\omega^2+4\tilde\kappa^2)
}{\(\omega^2-\tilde\kappa(2\kappa+\kappa_0)
\)^2+\omega^2\(2\tilde\kappa+\kappa+\kappa_0\)^2 }.\L{4.6}
\EY
These spectra are obtained for the arbitrary relations between the
frequency parameters $\kappa,\;\tilde\kappa$, and $\kappa_0$

Generally speaking, both spectra turn out to be dependent on the
parameter $\xi$, i. e., on the photon statistics exciting laser.

\subsection{Spectra in the {\it (2L)}-channel }
First of all, let us check what the photon statistics take place
from the isolated (without the three-level medium) {\it (2L)}. We
remember that photocurrent (\ref{4.6}) determines the photon
statistics in the presence of three-level lasing that can
seriously affects on the {\it (2L)}. In order to obtain the
required spectrum for the isolated laser, we must to put in
Eq.~(\ref{4.6}) first $\kappa_0=0$ and second $\tilde\kappa=0$. A
sense of the first requirement is very clear because just the
coefficient $\kappa_0$ determines the connection between lasers.
The second requirement is connected with equality (\ref{3.5}) that
as $\kappa_0=0$ survives only if $\tilde n=0$ or $\tilde\kappa=0$.
Because under the construction of the master equation (\ref{3.1})
in Ref.~\cite{Rost} we have employed the limit of the small photon
fluctuations or equivalently $\tilde n\gg1$.

So putting in (\ref{4.6}) $\tilde\kappa,\kappa=0$, the
photocurrent spectrum for the isolated {\it (2L)} reads:

\BY
&&(\delta i^2)_\omega/\o { i}=1+2\xi\frac{\kappa^2
}{\omega^2+\kappa^2 }.\L{4.7}
\EY
One can see that the photon statistics of the isolated {\it (2L)}
is dependent on the statistical Mandel parameter $\xi$ and can be
regulated by means of varying this parameter connected directly
with the statistical parameter $p$ of the {\it (2L)} pump.

Switching the {\it (3L)} on transforms the photon statistics of
the {\it (2L)}: the more $\kappa_0$, the more transformation. In
most interesting for us case as $\kappa_0\gg\kappa$ and
$\kappa=\tilde\kappa$ the generation turns out to be Poissonian
independently of $\xi$:
\BY
&& (\delta i^2)_\omega=\o { i}.
\EY
This situation is represented as quite natural because most of the
intracavity field in the {\it (2L)} turns out to be
non-controlled: most of the field is spent on the excitation of
the three-level medium and only small part leaves the cavity for
photodetecting.

\subsection{Spectra in the {\it (3L)}-channel: {\it induced photon statistics}}
Let us analyze  the photocurrent spectrum from the {\it (3L)} that
is given by (\ref{4.5}). We again can consider two interesting
limits, namely the high and low rate of the coherent excitation of
the three-level medium.

In the case as $\kappa_0\ll\kappa$, the photocurrent reads:
\BY
&&(\delta \tilde i^2)_\omega/\o {\tilde
i}=1-2\tilde\kappa^2\frac{\omega^2+\kappa^2
}{\(\omega^2-2\tilde\kappa\kappa
\)^2+\omega^2\(2\tilde\kappa+\kappa\)^2 }.\L{}
\EY
One can see that the spectrum turns out to be independent of
$\xi$. This perfectly corresponds to our ideas about the processes
taking place in the laser.  Really, only small part of the {\it
(2L)} radiation excites the three-level medium and this part is
able to be taken in by the medium only as the Poissonian field
independently of the real photon statistics in the {\it (2L)}.

One can see that, on the zero frequency, the noise is reduced
below the shot level  by halves: $ (\delta i^2)_{\omega=0}=\o {
i}/2$. This effect is well known and was discussed already in
detail Ref.~\cite{Ritsch}.

The opposite  limit $\kappa_0\gg\kappa$ is more interesting for us
here. Putting simultaneously  $\kappa=\tilde\kappa$, one can
obtain:
\BY
&& (\delta \tilde i^2)_\omega/\o { \tilde
i}=1+2\xi\frac{\kappa^2(\omega^2+\kappa_0^2)}{\(\omega^2-\kappa\kappa_0
\)^2+\omega^2\kappa_0^2 }\approx
1+2\xi\frac{\kappa^2}{\omega^2+\kappa^2 } .\L{}
\EY
The last equality can be obtained, if we take into account that
$\omega^2<\kappa^2\ll\kappa\kappa_0\ll\kappa_0^2$. So the observed
photocurrent in the {\it (3L)}-channel is of the same as for  the
isolated {\it (2L)} (\ref{4.7}). We can conclude that the {\it
(3L)} photon statistics duplicate the photon statistics of the
coherently exciting laser. We call this phenomenon the {\it
induced photon statistics (IPS) }.

%\newpage

\section{Excitation of the {\it (3L)} by radiation
from the auxiliary Poissonian laser involved into the feedback
process \L{V}}
The { \it IPS } is not only the fundamental but can carry some
auxiliary functions out. For example, this can be employed for
studying the laser photon statistics, when a radiation from the
laser is inaccessible for a direct observation. Just this
situation takes place for the laser that is included into the {\it
FBL}. We remember that in the effective feedback, the current
noises turn out to be reduced below the quantum limit. A lot of
persons, who work in this area, believe that, in this case, they
deal with  the sub-Poissonian photon statistics in lasing. From
their standpoint, there is an only problem to find a way for the
use of this non-classical light.

At the same time according to Ref.~\cite{Milburn} the opposite
standpoint is soon more truthful. Although the feedback leads to a
stabilization of the photocurrent and reduction of the electron
noises below the quantum limit, nevertheless  this does not
concern to the corresponding light beam. The {\it FBL} in itself
is unable to produce any non-classical effects in the field. We
want to confirm this standpoint in the mental experiment with
employing the discussed above {\it ISP} phenomenon.

In this section, we shall try to answer question, what happens
with the photon statistics when the Poissonian laser is involved
into the feedback process. For that, we have to develop the theory
like the theory in the previous sections, taking into account
supplementary factor, namely the feedback (see Fig.~1).

The photocurrent stabilization in the FBL is achieved in the
following way: any positive (negative) fluctuation of the
photocurrent $\delta i=i-\o i$ is accompanied by the appropriate
reduction (increase) of the laser power. Physically this can be
established via the pump process. In order to take into account
the feedback, usually some phenomenological treatment is applied.
The simplest way for us is, in the basic laser equations, to
replace the mean pump rate $R$  on another value $\tilde R$ that
depends properly on the current fluctuations. For example, this
reads:
\BY
&&\tilde R= R(1-\lambda\; \delta i/\o i)\L{}.
\EY
The value $\lambda$  determines an efficiency of the feedback and
 is usually called the {\it feedback strength}.

Evidently, this phenomenological model is extremely simplistic,
because this does not take into account the time delays
characteristic of any similar scheme. However, for our qualitative
conclusions, similar details are not  important.

Generalizing the theory developed in Ref.~\cite{Rost} to the case
of the feedback presence, we can obtain the following equations
instead of
 (\ref{3.15}) and (\ref{3.16}):
 \BY
&&-i\omega\varepsilon_\omega=-(\kappa+\kappa_0)(1+\lambda)\varepsilon_\omega+
\tilde\kappa
\tilde\varepsilon_{\omega}+F_\omega-\lambda(1+\kappa_0/\kappa)S_\omega,\L{5.1}\\
&&-i\omega\tilde\varepsilon_{\omega}=-2\tilde\kappa\tilde\varepsilon_{\omega}+
\kappa_0\varepsilon_\omega+\tilde F_{\omega}\L{5.2}.
\EY
One can see that the feedback leads to an important transformation
in the laser equations, namely  the stochastic source $S_\omega$
connected with photodetecting occurs there.

Now we have everything to derive the photocurrent spectra in both
channels from the {\it (2L)} and the {\it (3L)}. First of all it
would be nice to be convinced that the proposed model of the
feedback works and  as a matter of fact the photocurrent from the
isolated {\it (2L)} turns out to be stabilized. For that we have
to put in Eqs.~ (\ref{5.2}) and
(\ref{5.3})$\kappa_0=\tilde\kappa=0$. Under this conditions, the
algebraic system is easy solved and after that it is not difficult
to calculate the required spectrum:
\BY
&&(\delta i^2)_\omega/\o
i=1-\kappa^2\frac{(1+\lambda)^2-1}{\omega^2+(1+\lambda)^2\kappa^2}.\L{5.4}
\EY
One can see as $\lambda\gg1$ the current noise turns out to be
perfectly reduced on the zero frequency. So our simplistic
feedback mechanism works properly.

 Switching the three-level laser
on transforms this formula to:
\BY
&&(\delta i^2)_\omega/\o{ i}=1+\frac{\kappa^2}{\kappa_0^2}
\;\frac{-2\kappa^2+4\lambda^2\kappa_0^2}{
\omega^2(1+\lambda)^2+\kappa^2(1+2\lambda)^2}.\L{5.5}
 \EY
Here again we have chosen the conditions  $\kappa_0\gg\kappa$ and
$\kappa=\tilde\kappa$. This is most interesting for us because
this choice ensures existing the {\it ISP} phenomenon. It is not
difficult to see that now the electron statistics under detecting
the {\it (2L)} radiation turns out to be super-Poissonian. The
last is important for our understanding the situation. The
 switching the three-level laser on leads to the essential
 perturbation  of the {\it (2L)} state.
 This means that  our measuring procedure is  by
 no means non-demolition one.

According to our consideration here the electron statistics in the
{\it (3L)}-channel provides us with a possibility to make the
correct conclusion relative to the photon statistics not only in
the {\it (3L)}-channel but in the {\it (2L)}-channel too.
Certainly, for that we need to guarantee existing {\it ISP}
phenomenon, then the spectrum reads:
\BY
&&(\delta\tilde i^2)_\omega/\o{\tilde i}=1-\frac{\kappa}{\kappa_0}
\;\frac{2\kappa^2+2\lambda\kappa\kappa_0-\lambda^2\kappa_0^2}{
\omega^2(1+\lambda)^2+\kappa^2(1+2\lambda)^2}.\L{}
 \EY
For the high effective feedback $\lambda\gg1$ the spectrum is
given by:
\BY
&&(\delta\tilde i^2)_\omega/\o{\tilde i}=1+
\;\frac{\kappa\kappa_0}{ \omega^2+4\kappa^2}.\L{}
\EY
One can see that the photon statistics in the {\it (2L)}-channel
turn out to be strongly super-Poissonian, although the
corresponding the electron statistics are sub-Poissonian
(\ref{5.4}). Certainly, this is in the qualitative agreement with
Ref.~\cite{Milburn}.

\section{Conclusion\L{VI}}
We have considered the phenomenon in  three-level lasing when the
photon statistics of a generation  duplicate the photon statistics
of another laser that pumps the first coherently. We called this
the {\it induced photon statistics ({\it ISP})}.

We have discussed an applied possibility of this phenomenon on the
example of the laser involved into the feedback process.  Because,
in this case, it is impossible the direct observation of the laser
beam, we suggested to use for analysis the {\it ISP} by means of
applying the investigated laser for the coherent pump of the
auxiliary three-level laser. As a result, we have make a
conclusion that the photon statistics of the laser in the feedback
loop turn out to be strongly super-Poissonian although the
corresponding the photocurrent has the  noises reduced effectively
on the zero frequency below the quantum limit.

\section{Acknowledgement\L{VII}}

This work was performed within the Franco-Russian cooperation
program "Lasers and Advanced Optical Information Technologies"
with financial support from the following organizations: INTAS
(Grant No. 01-2097)and RFBR (Grant No. 05-02-19646).

\newpage

 \begin{figure}[t]
% \epsfxsize=12cm
% \centerline{\epsfbox{Fig2.eps}}
 \includegraphics[width=120mm]{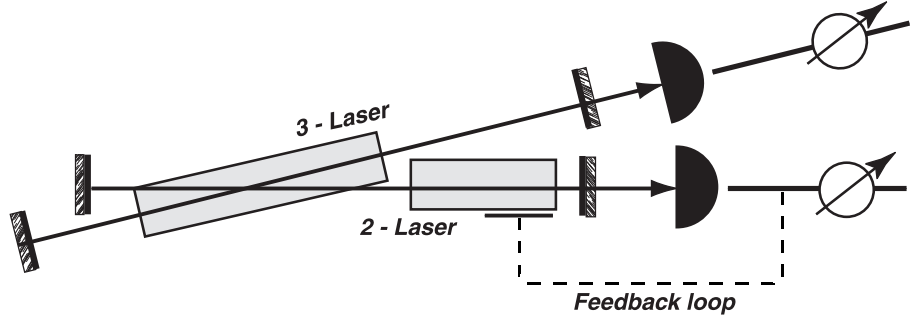}
 \caption{Mental experimental setup with two coupled lasers:
 '2-laser' is the two-level laser included into the feedback loop;
  '3-laser' is the three-level laser }
% \label{fig2}
 \end{figure}

 \begin{figure}[t]
% \epsfxsize=12cm
% \centerline{\epsfbox{fig3.eps}}
 \includegraphics[width=120mm]{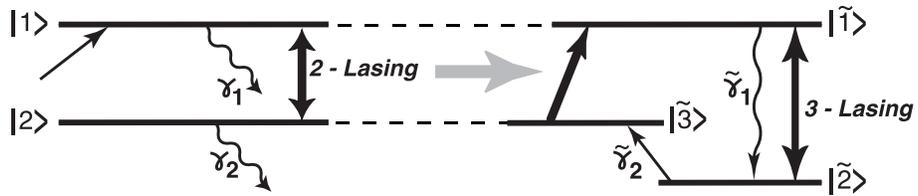}
 \caption{Atomic energetic configurations for the 2-laser (on the left) and the 3-laser
 (on the right) }
% \label{fig3}
 \end{figure}

\end{document}